\documentclass[a4paper,11pt]{article}
\usepackage{pos}
\usepackage{lineno}

\graphicspath{
    {IMAGES/}
}

\title{Characterizing system dynamics with
two-particle transverse momentum correlations
in pp collisions at $\sqrt{s} = 7\;\text{TeV}$ and
p--Pb collisions at $\sqrt{s_{\rm NN}} = 5.02\;\text{TeV}$}
\ShortTitle{Two-particle transverse momentum correlations}

\author*[a,1]{Víctor González}

\affiliation[a]{Department of Physics and Astronomy, Wayne State University, W. Hancock 666, Detroit, 48201, USA}

\note{For the ALICE Collaboration}
\emailAdd{victor.gonzalez@cern.ch}

\abstract{The two-particle differential transverse momentum correlator $G_{2}$ recently measured in Pb--Pb collisions, emerged as a powerful tool to gain insight into particle production mechanisms and to infer transport properties such as the ratio of shear viscosity to entropy density of the medium created in Pb--Pb collisions.
In this poster, recent ALICE measurements of this correlator in pp collisions at $\sqrt{s}=7$ and p--Pb collisions at $\sqrt{s_{\rm NN}}=5.02$ TeV are presented to search, in particular, for viscous effects expected to arise in fluid-like systems produced in these collisions. The strength and shape of the correlator are studied as a function of produced particle multiplicity to look for longitudinal broadening that might reveal the presence of viscous effects in these smaller systems. The measured correlators and their evolution from pp and p–Pb to Pb–Pb are additionally compared to predictions from Monte Carlo models, and the potential presence of viscous effects is discussed.}

\FullConference{%
  41st International Conference on High Energy physics - ICHEP2022\\
  6-13 July, 2022\\
  Bologna, Italy
}

\begin{document}
\maketitle

\section{Introduction}

The specific shear viscosity, $\eta/s$, is the transport coefficient that most directly controls the expansion of viscous systems. Recent measurements of the longitudinal broadening of the $G_{2}$ correlator~\cite{Gavin:2006xd} by the ALICE Collaboration~\cite{ALICE:2019smr} yield a QGP $\eta/s$ range compatible with estimations based on anisotropic flow. This work measures the same correlator in pp and p--Pb collisions seeking for evidence of longitudinal broadening that may indicate the presence of viscous effects. 
\begin{figure}[b!]
  \centering
  \includegraphics[width=0.32\textwidth,keepaspectratio=true,clip=true,trim=0pt 0pt 0pt 0pt]
  {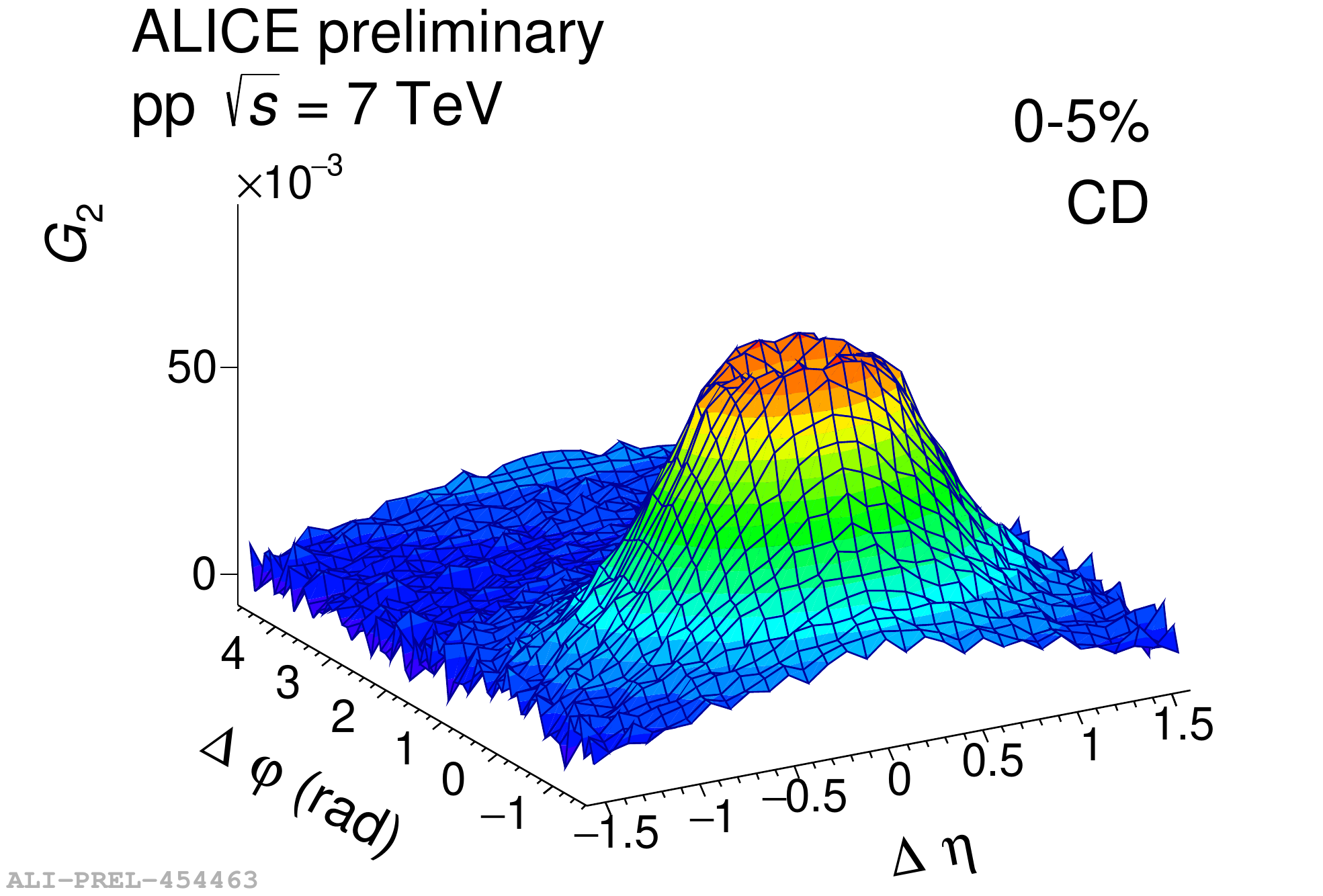}
  \includegraphics[width=0.32\textwidth,keepaspectratio=true,clip=true,trim=0pt 0pt 0pt 0pt]
  {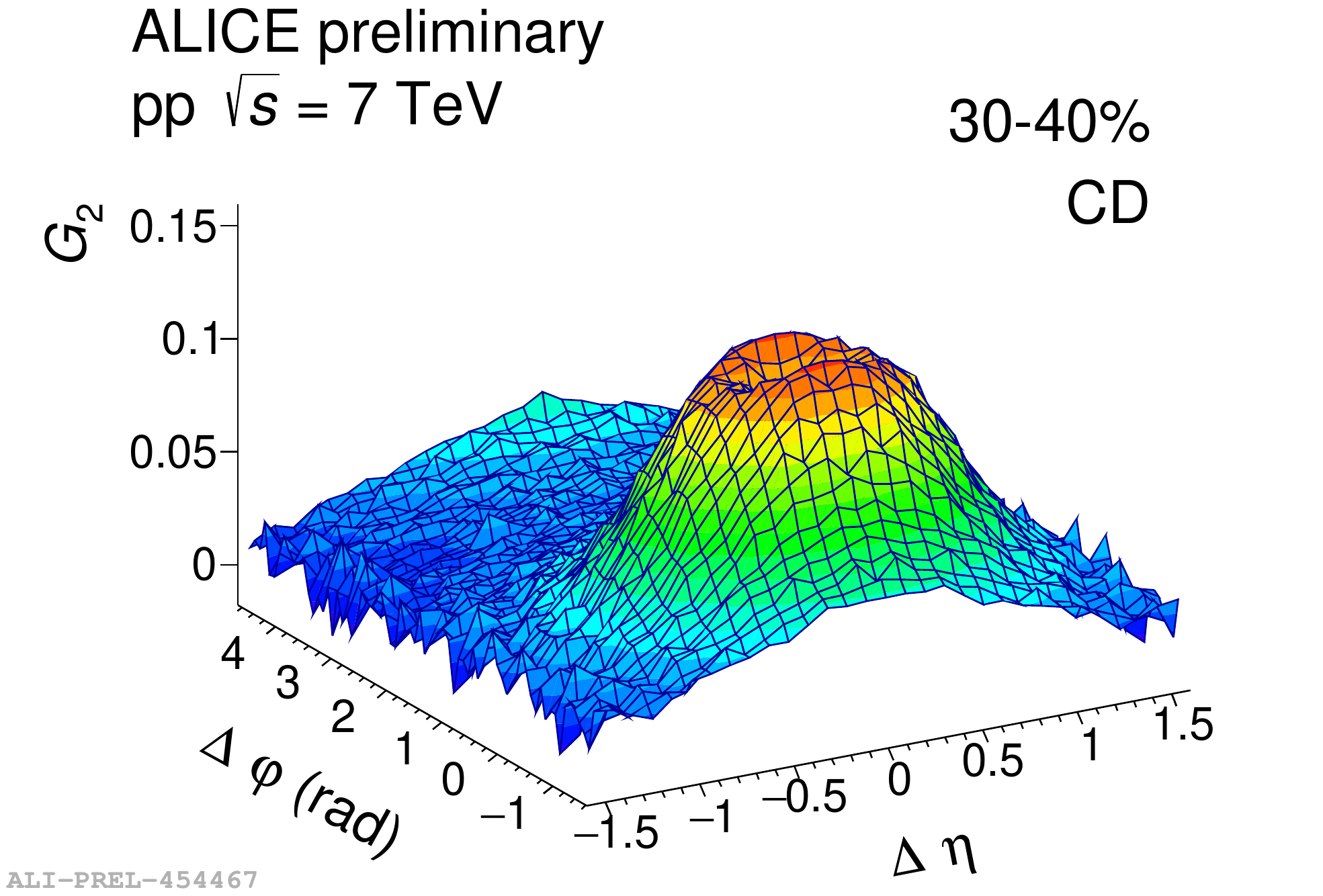}
  \includegraphics[width=0.32\textwidth,keepaspectratio=true,clip=true,trim=0pt 0pt 0pt 0pt]
  {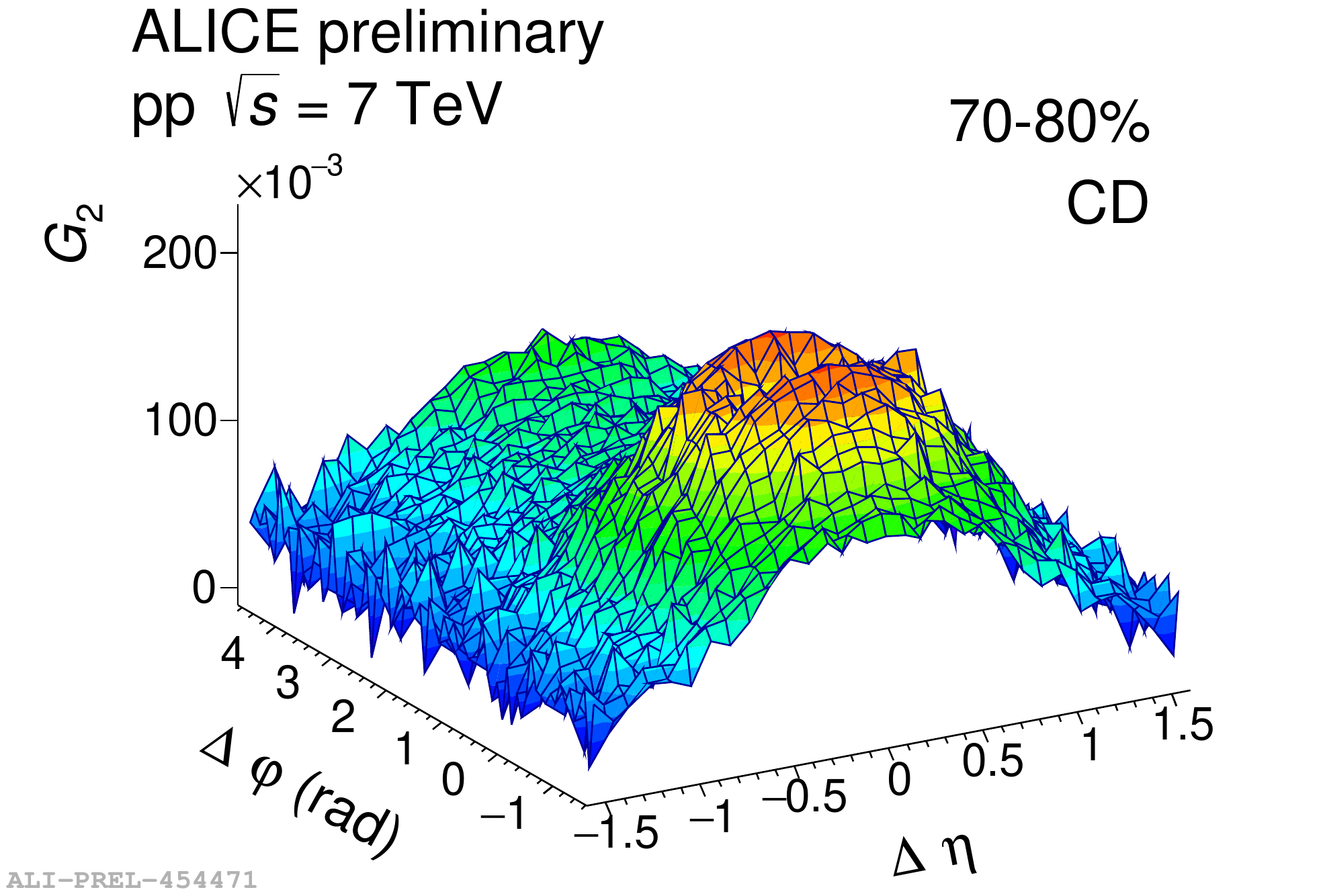}\\
  \includegraphics[width=0.32\textwidth,keepaspectratio=true,clip=true,trim=0pt 0pt 0pt 0pt]
  {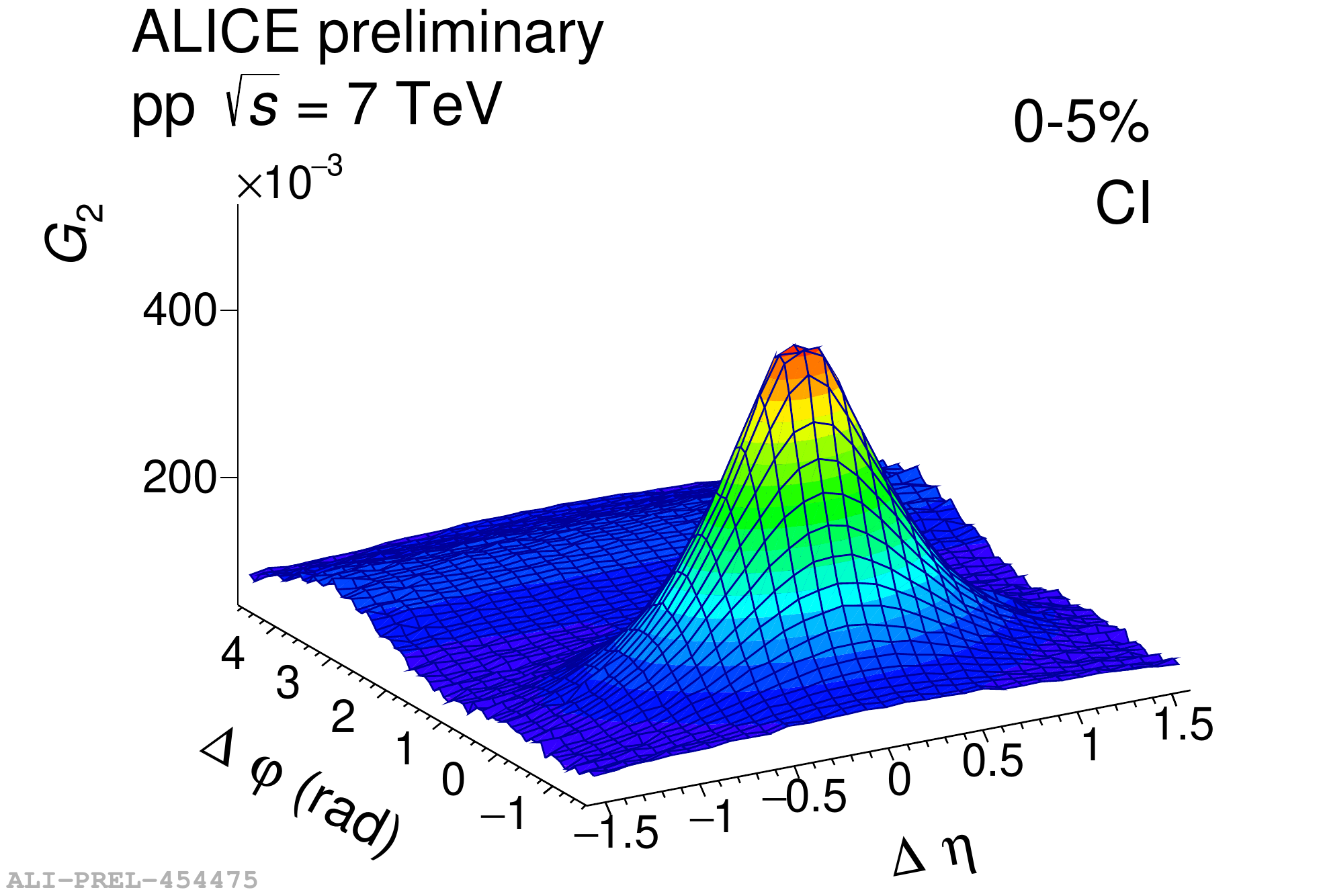}
  \includegraphics[width=0.32\textwidth,keepaspectratio=true,clip=true,trim=0pt 0pt 0pt 0pt]
  {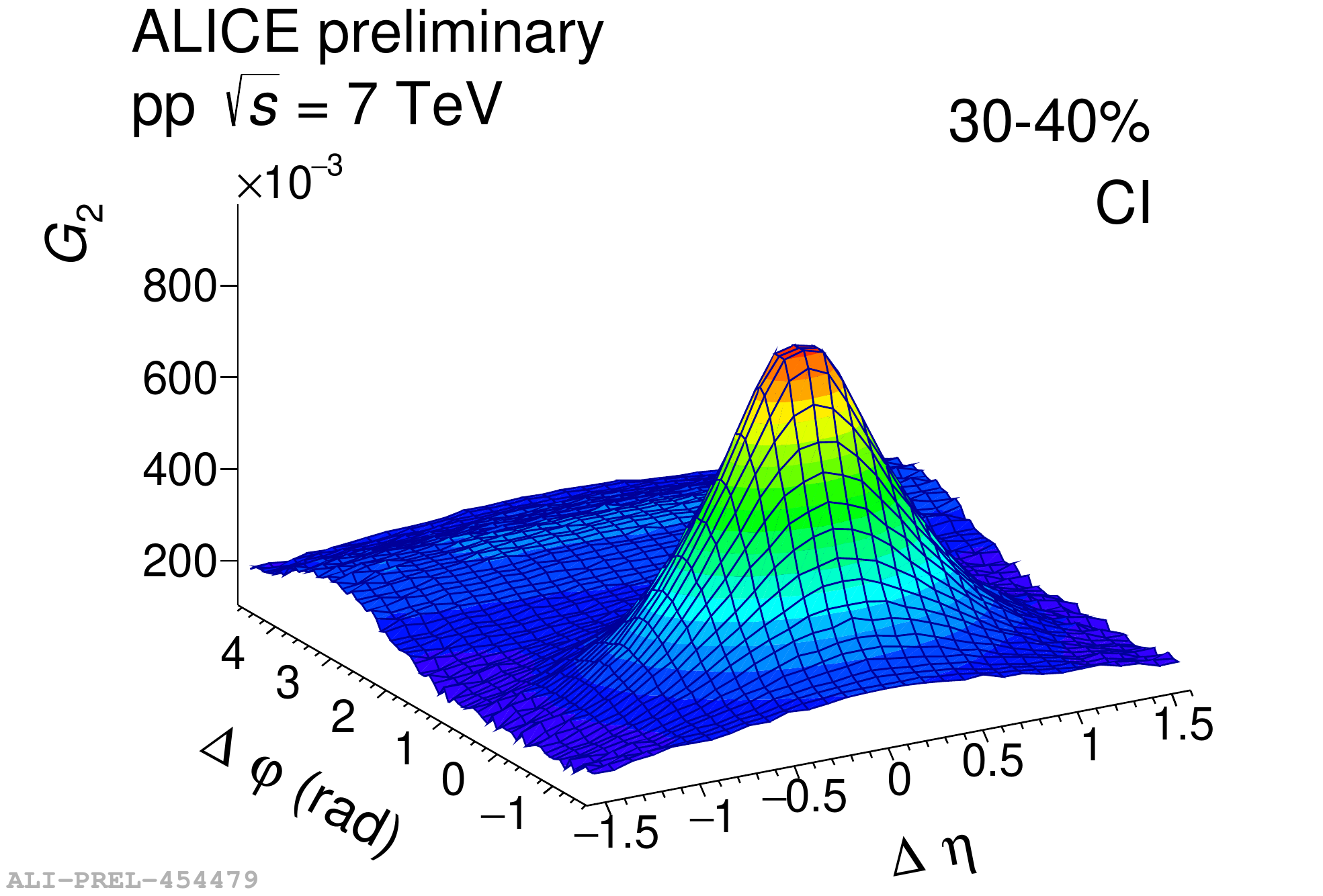}
  \includegraphics[width=0.32\textwidth,keepaspectratio=true,clip=true,trim=0pt 0pt 0pt 0pt]
  {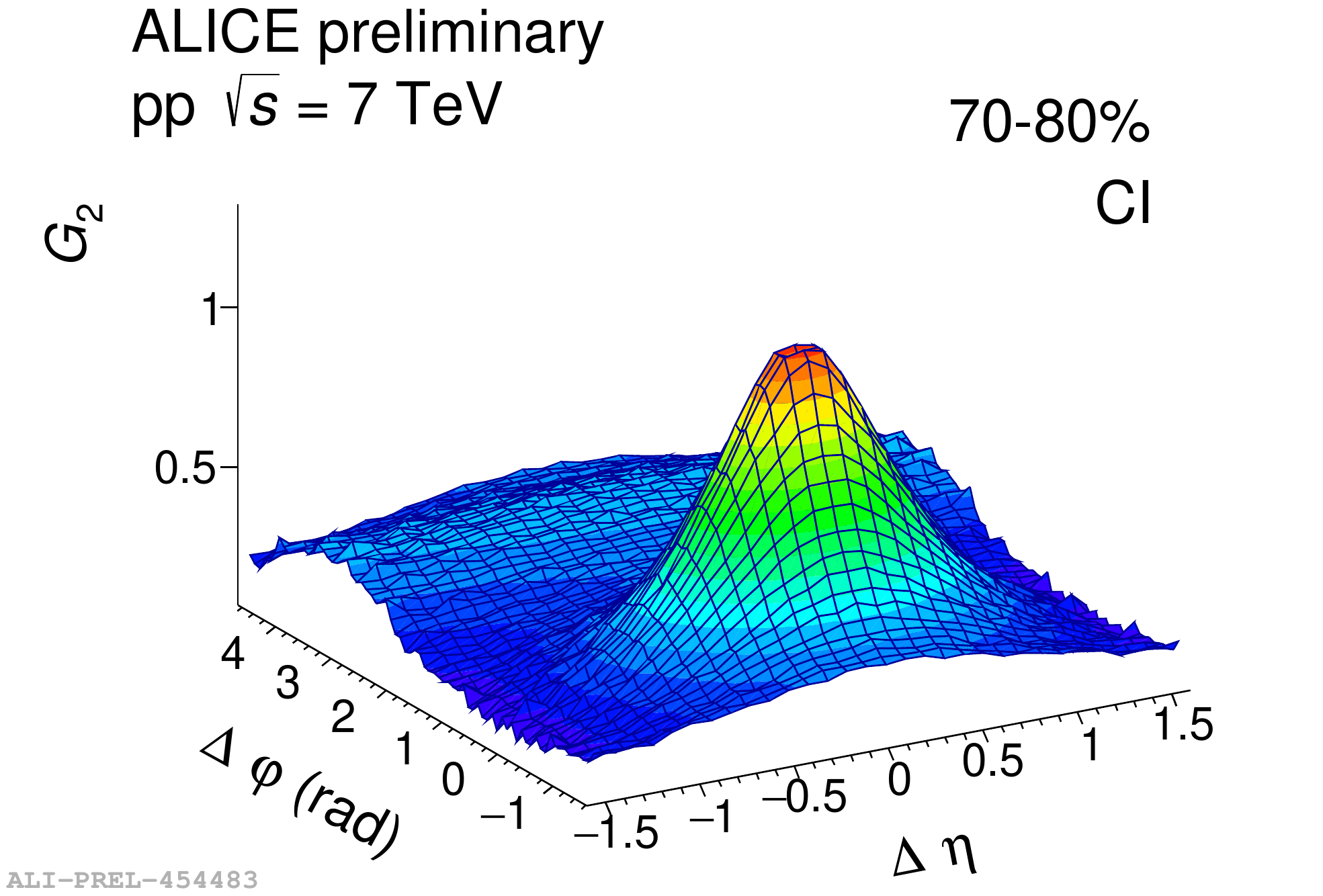}
  \caption{Two-particle transverse momentum correlators $G_{2}^{\rm CD}$ (top row) and $G_{2}^{\rm CI}$ (bottom row) for selected multiplicity classes in pp collisions at $\sqrt{s}=7\;\text{TeV}$.}
  \label{fig:pp2dcdci}
\end{figure}
\section{Experimental measurements}
\begin{figure}[hb]
\centering
  \includegraphics[width=0.245\textwidth,keepaspectratio=true,clip=true,trim=0pt 0pt 30pt 0pt]
  {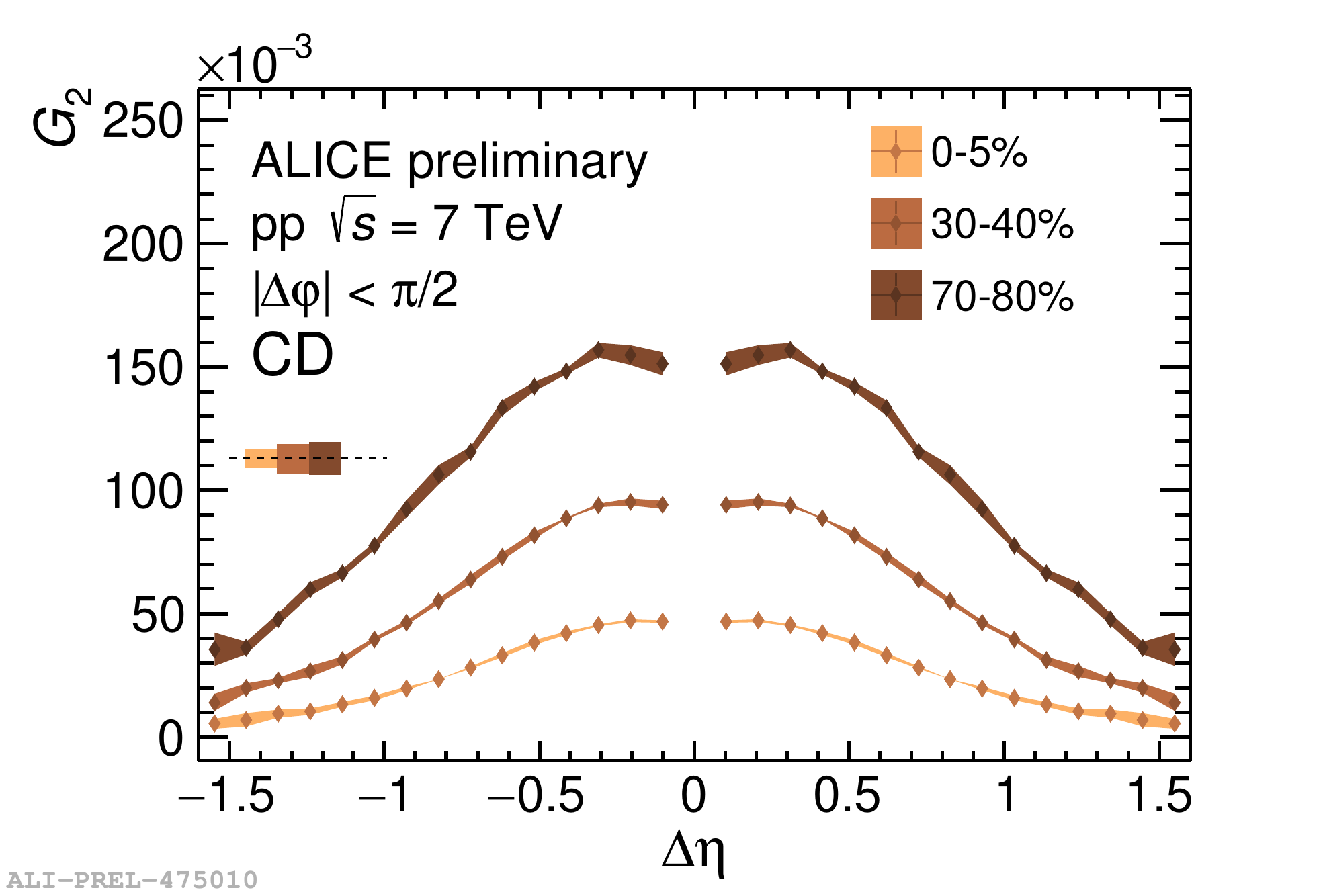}
  \includegraphics[width=0.245\textwidth,keepaspectratio=true,clip=true,trim=0pt 0pt 30pt 0pt]
  {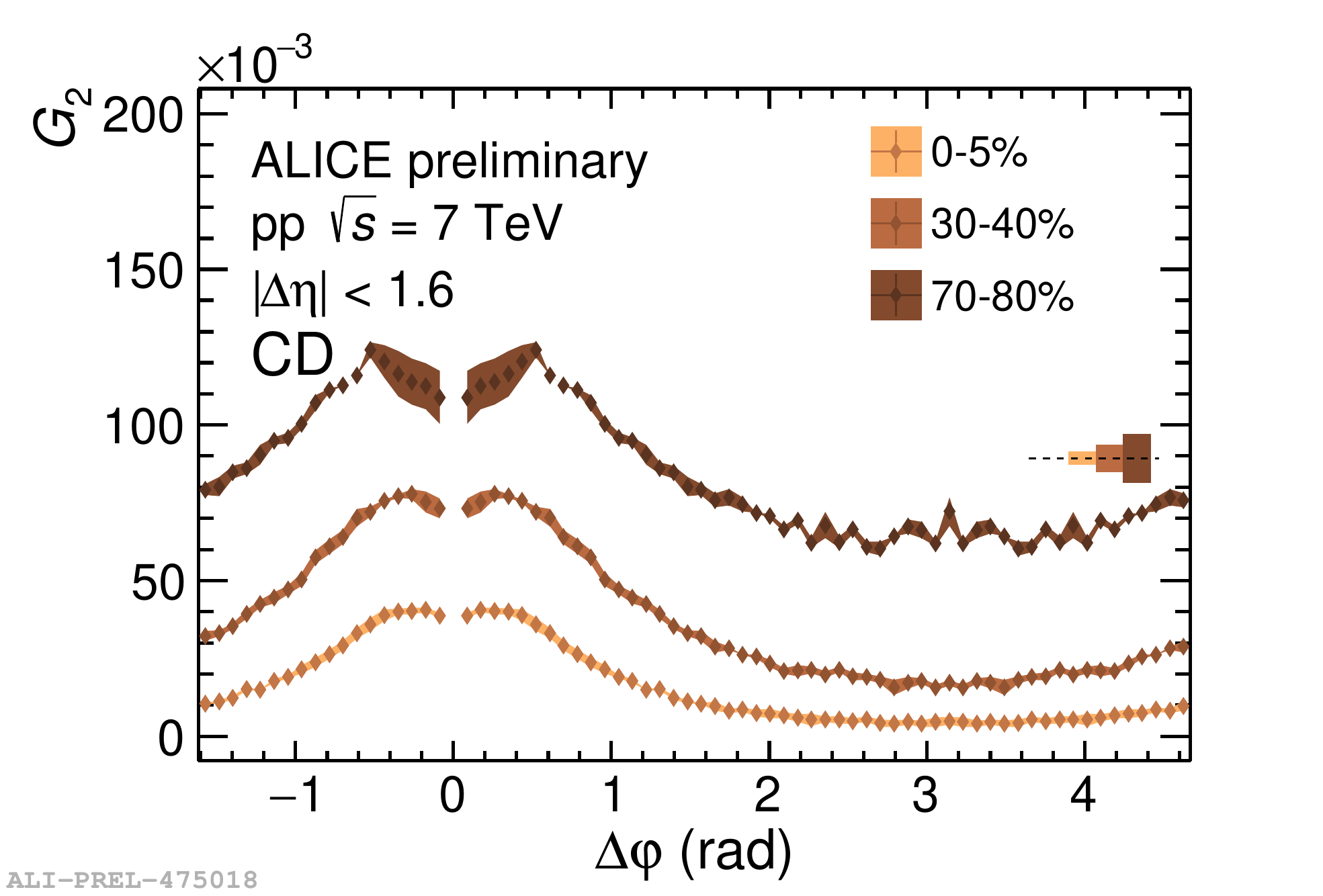}
  \includegraphics[width=0.245\textwidth,keepaspectratio=true,clip=true,trim=0pt 0pt 30pt 0pt]
  {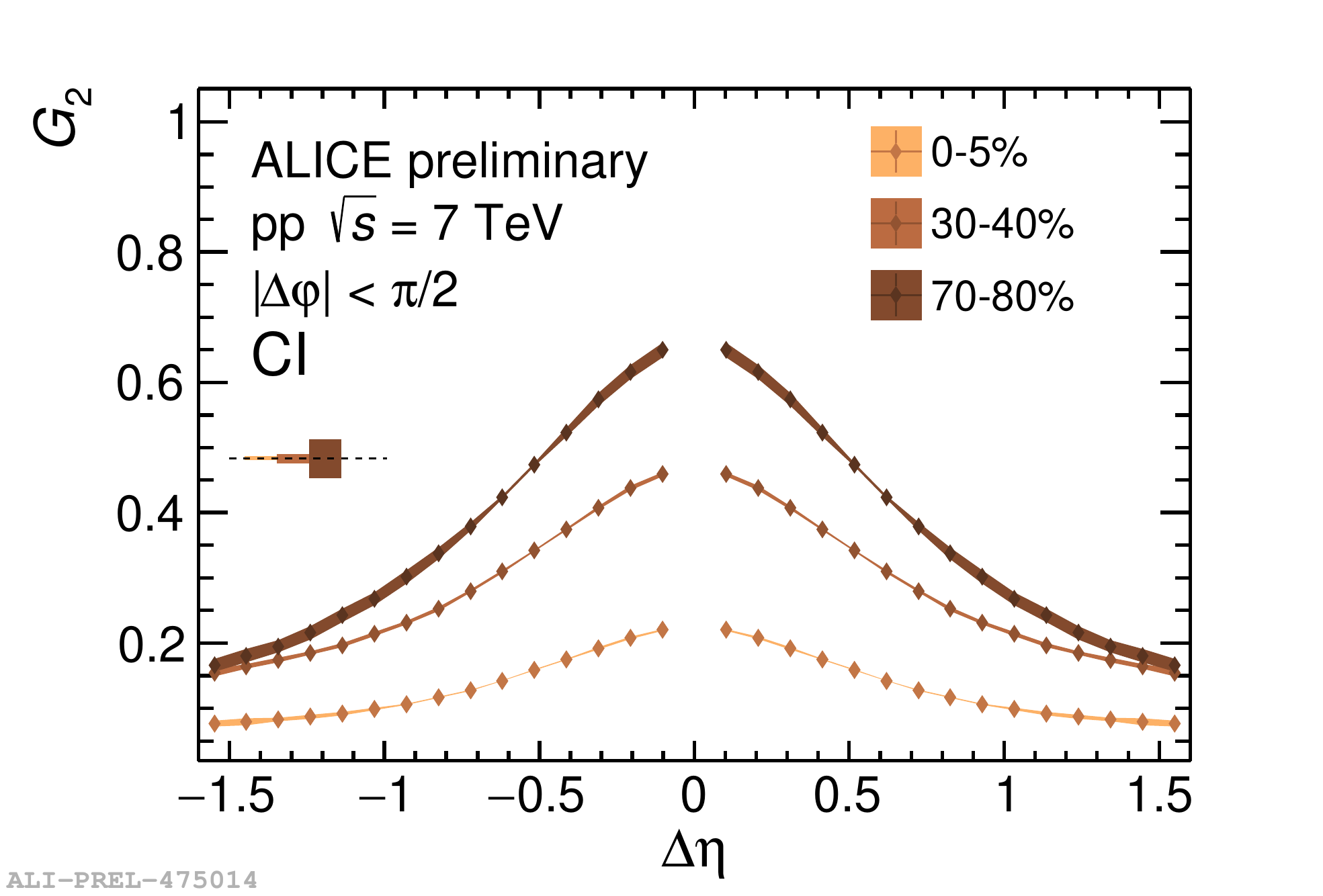}
  \includegraphics[width=0.245\textwidth,keepaspectratio=true,clip=true,trim=0pt 0pt 30pt 0pt]
  {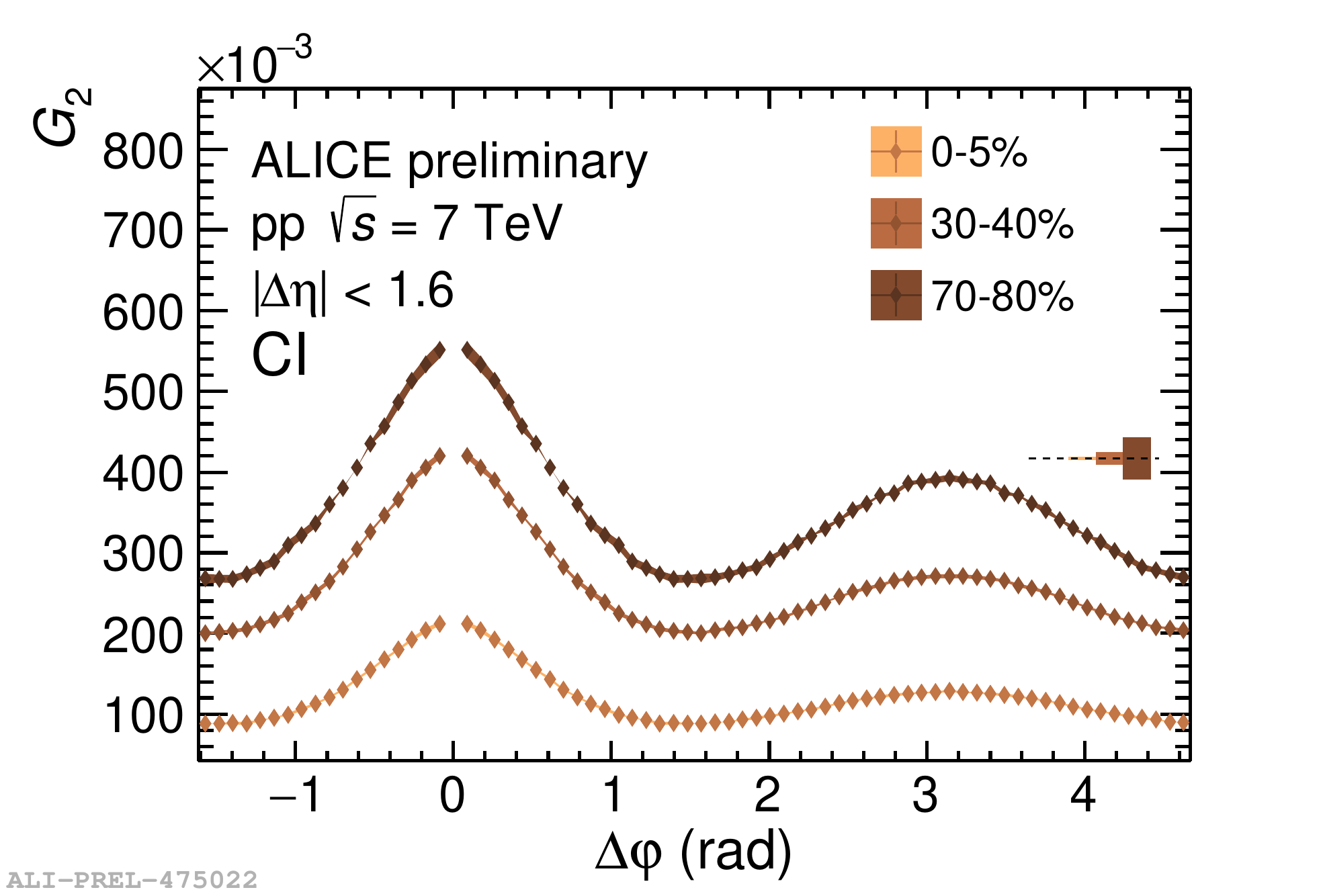}\\
  \includegraphics[width=0.245\textwidth,keepaspectratio=true,clip=true,trim=0pt 0pt 30pt 0pt]
  {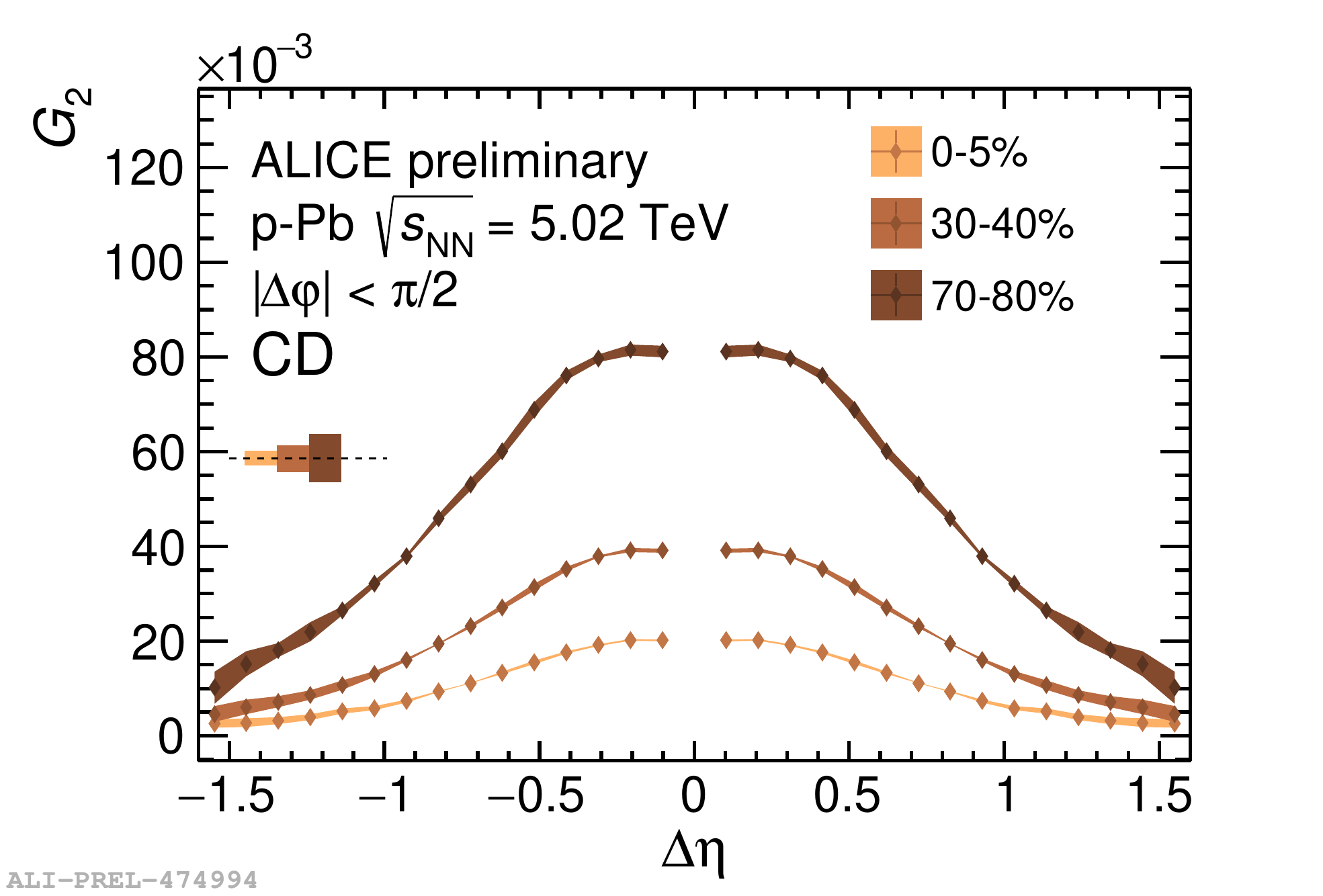}
  \includegraphics[width=0.245\textwidth,keepaspectratio=true,clip=true,trim=0pt 0pt 30pt 0pt]
  {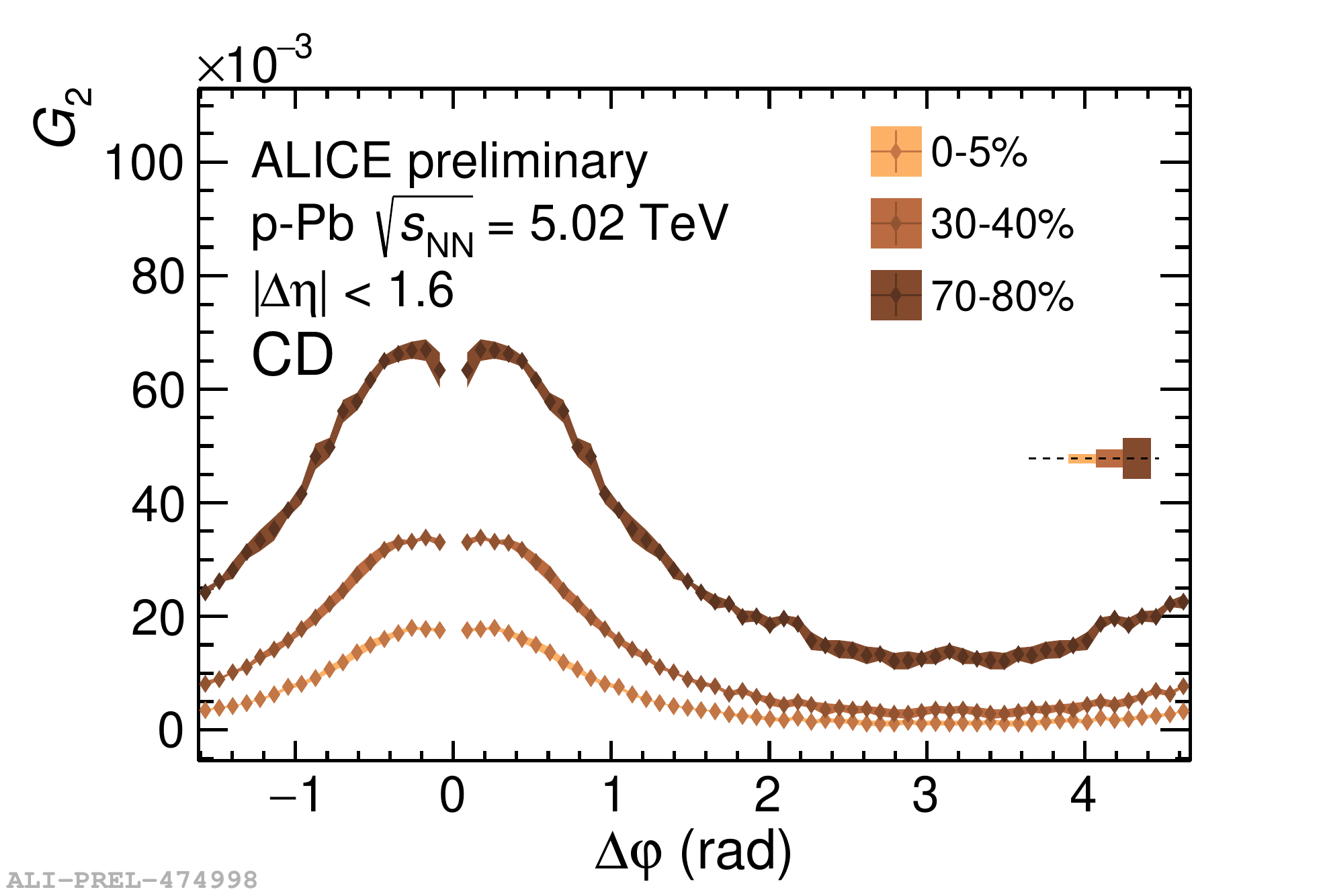}
  \includegraphics[width=0.245\textwidth,keepaspectratio=true,clip=true,trim=0pt 0pt 30pt 0pt]
  {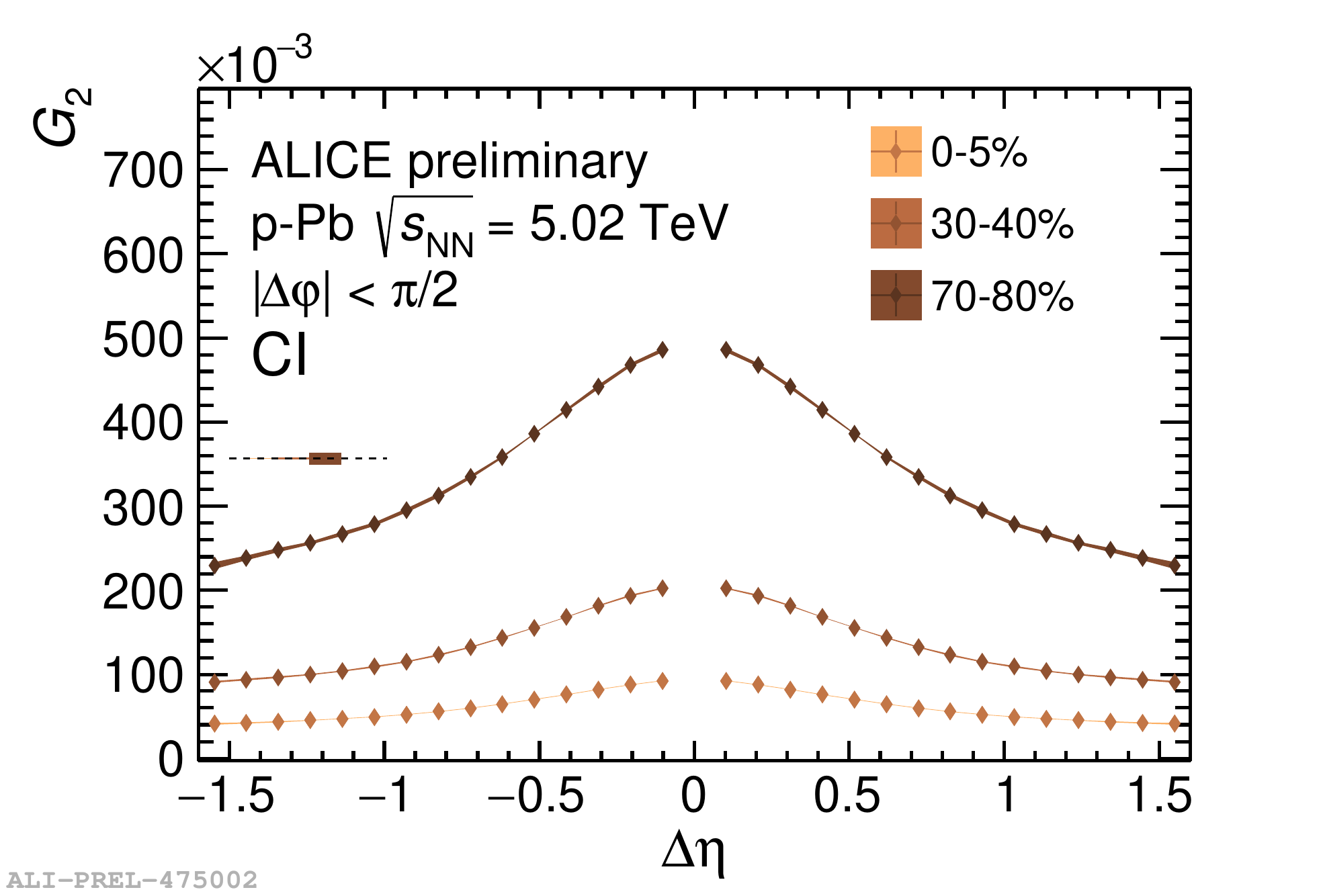}
  \includegraphics[width=0.245\textwidth,keepaspectratio=true,clip=true,trim=0pt 0pt 30pt 0pt]
  {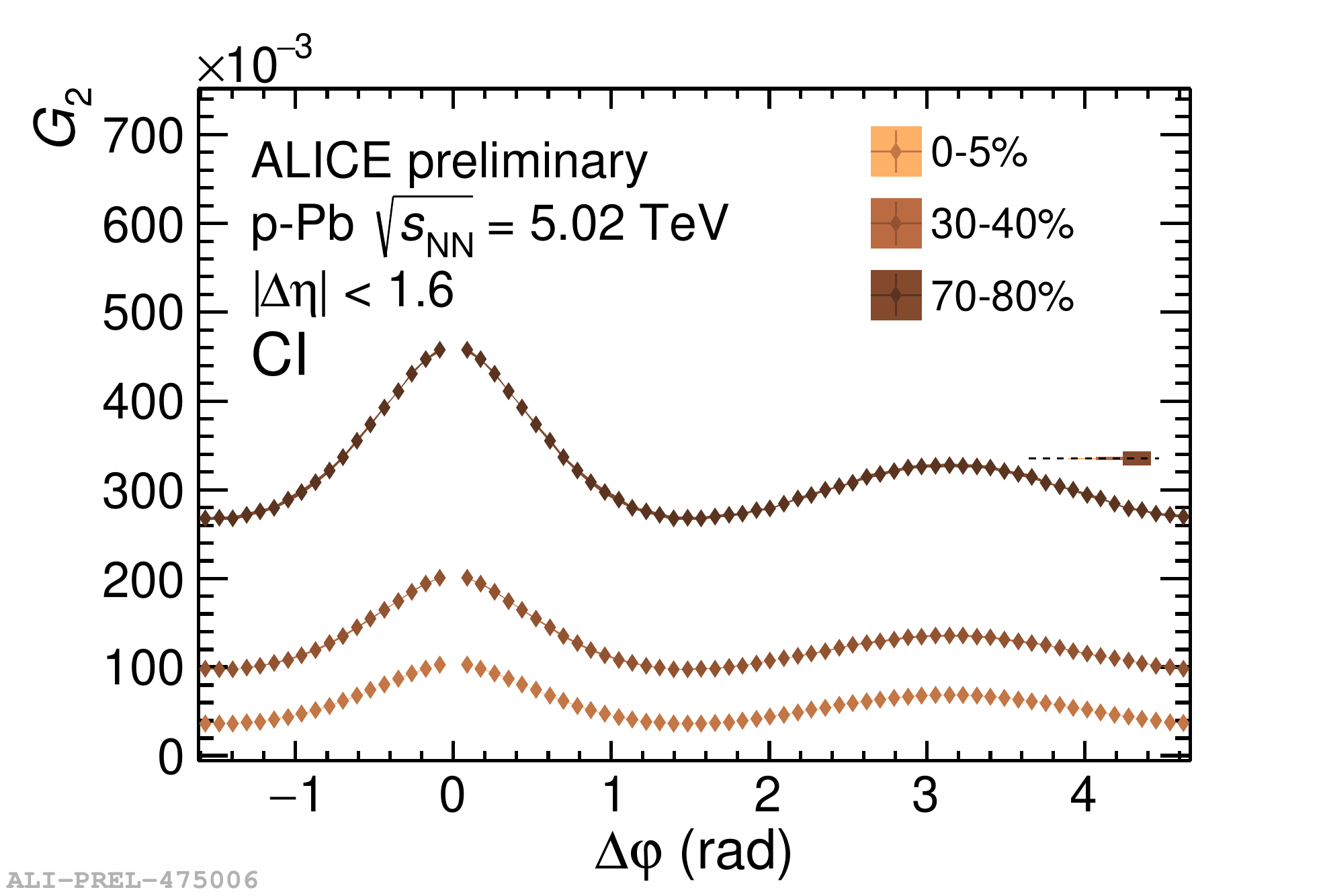}
\caption{Two-particle transverse momentum correlators $G_{2}^{\rm CD}$ (two left columns) and $G_{2}^{\rm CI}$ (two right columns) longitudinal (left) and azimuthal (right) projections for selected multiplicity classes in pp collisions at $\sqrt{s}=7\;\text{TeV}$ (top row) and p--Pb collisions at $\sqrt{s_{\rm NN}}=5.02\;\text{TeV}$ (bottom row).}
\label{fig:ppppbproj}
\end{figure}

The presented results correspond to measurements based on $6.4\times10^7$ selected minimum bias (MB) pp collisions at center-of-mass energy $\sqrt{s} = 7\;\text{TeV}$ and $5.4\times10^7$ MB p--Pb collisions at center-of-mass energy per nucleon--nucleon collision $\sqrt{s_{\rm NN}} = 5.02\;\text{TeV}$ recorded with the ALICE detector in the years 2010 and 2013, respectively~\cite{1748-0221-3-08-S08002}.
Results are reported in nine multiplicity classes corresponding to 0--5\% (highest multiplicity), 5--10\%, ..., 70--80\% (lowest multiplicity) of the total interaction cross section. The analysis uses tracks of charged particles reconstructed in the kinematic acceptance of $0.2\leq p_{\rm T} \leq 2.0\;\text{GeV}/c$ and  $|\eta|<0.8$.
\begin{figure}[ht]
  \centering
  \includegraphics[width=0.32\textwidth,keepaspectratio=true,clip=true,trim=0pt 0pt 0pt 0pt]
  {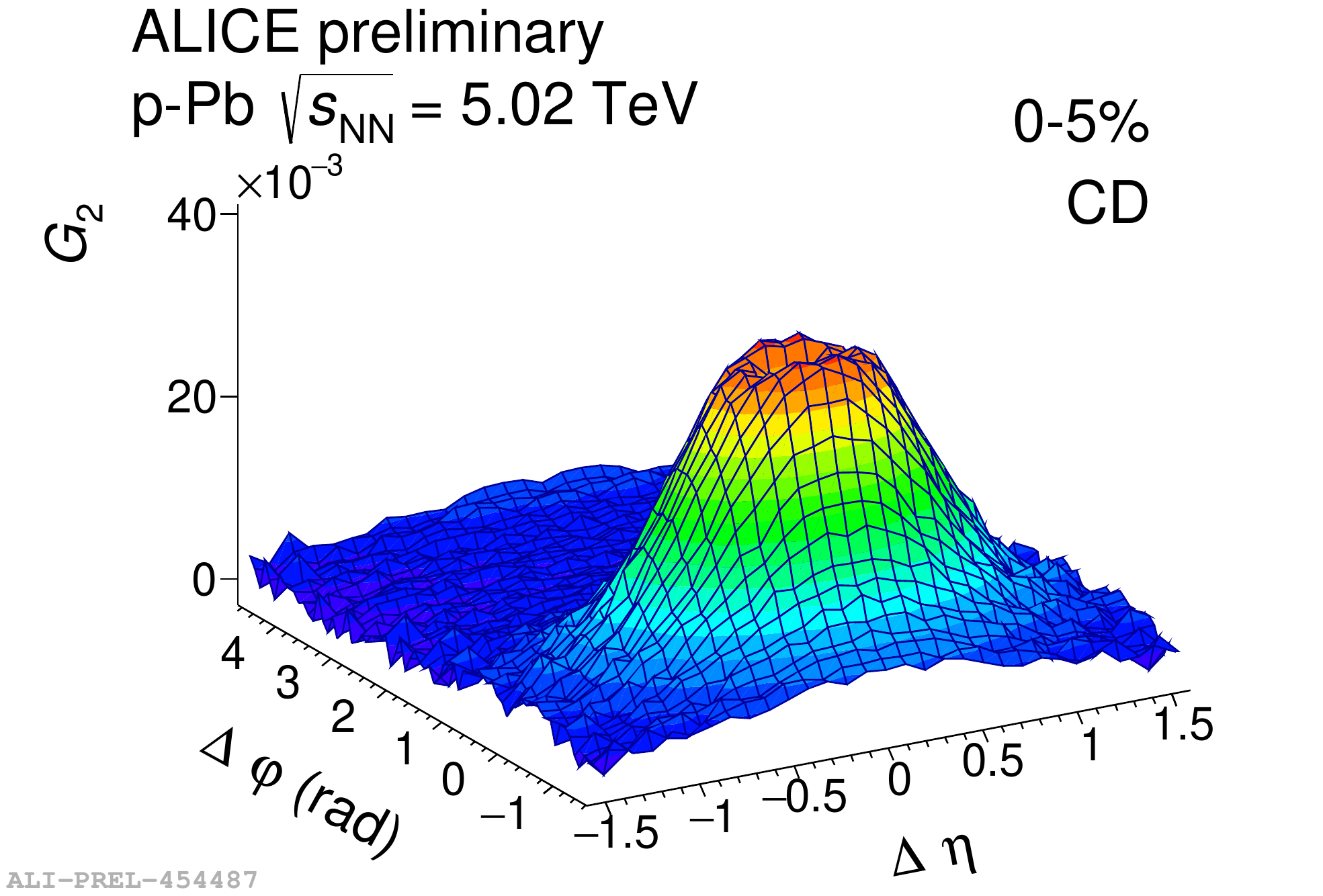}
  \includegraphics[width=0.32\textwidth,keepaspectratio=true,clip=true,trim=0pt 0pt 0pt 0pt]
  {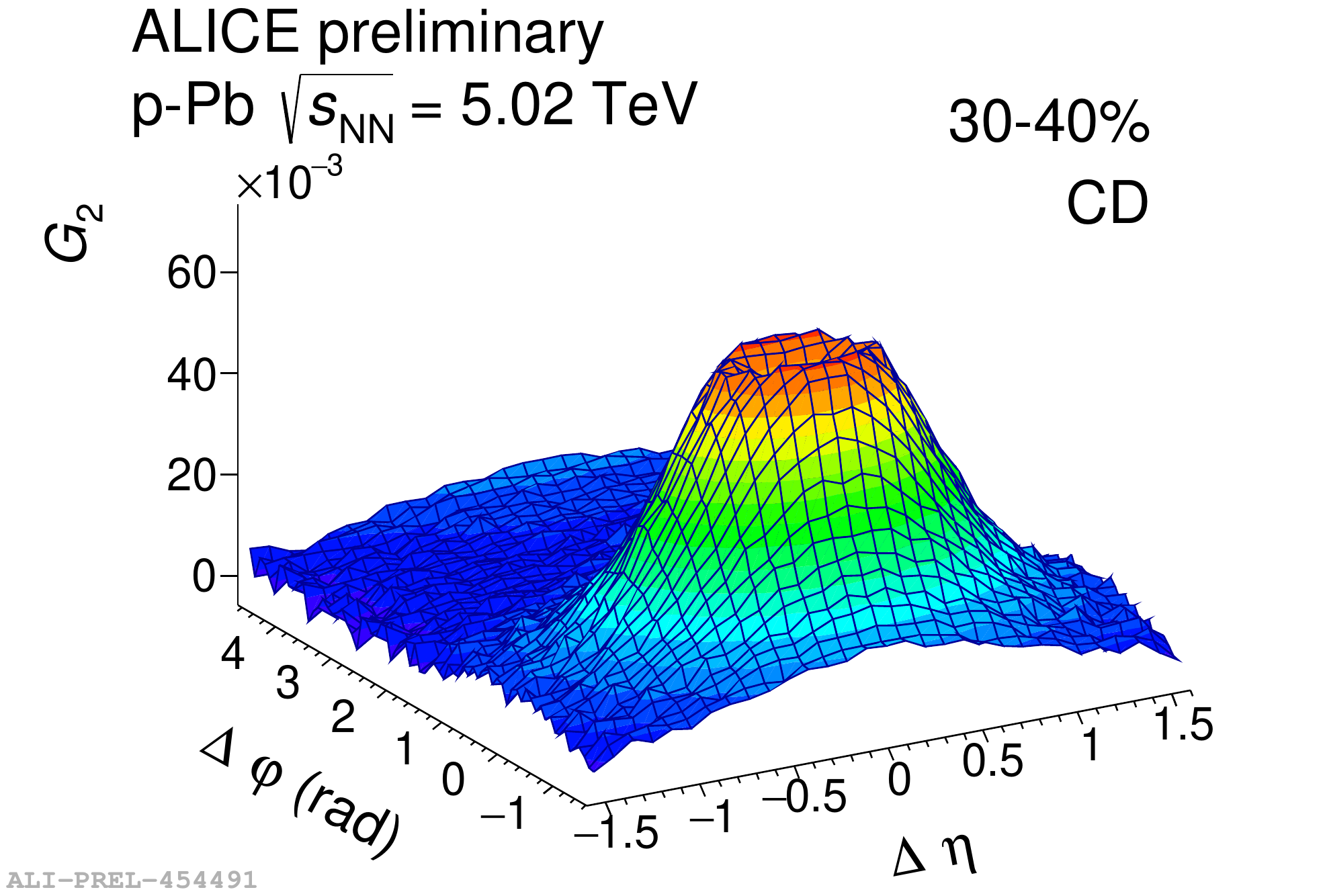}
  \includegraphics[width=0.32\textwidth,keepaspectratio=true,clip=true,trim=0pt 0pt 0pt 0pt]
  {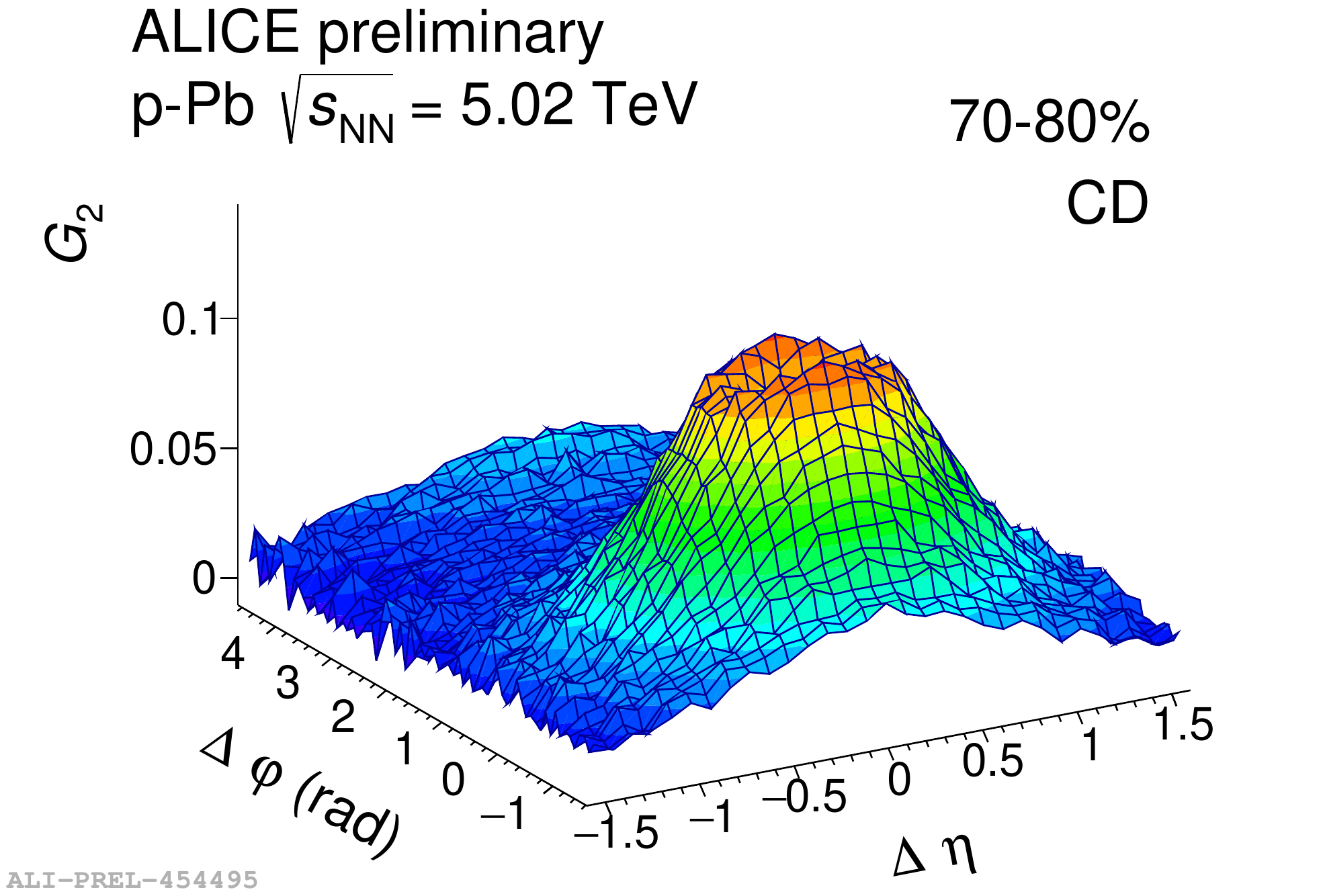}\\
  \includegraphics[width=0.32\textwidth,keepaspectratio=true,clip=true,trim=0pt 0pt 0pt 0pt]
  {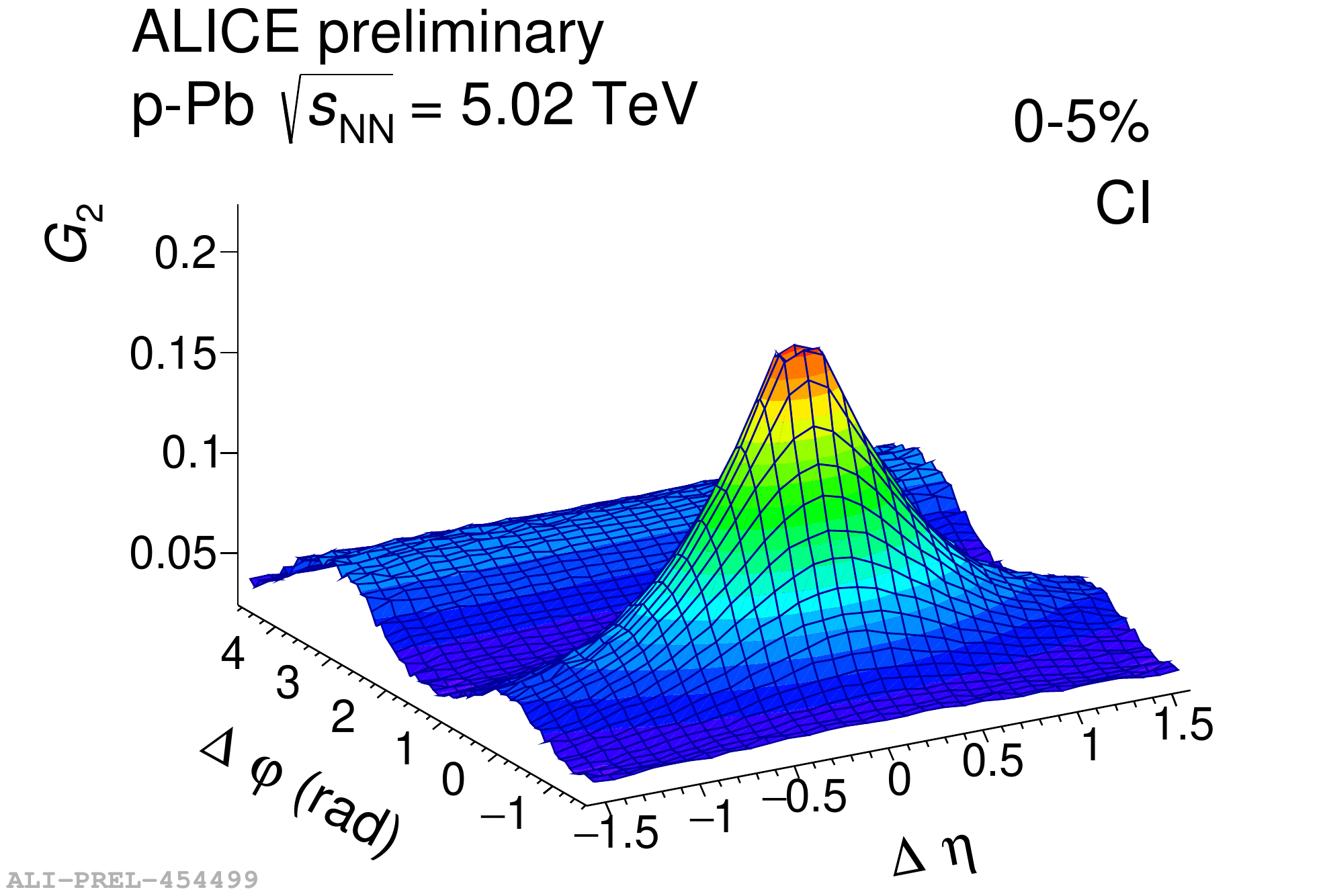}
  \includegraphics[width=0.32\textwidth,keepaspectratio=true,clip=true,trim=0pt 0pt 0pt 0pt]
  {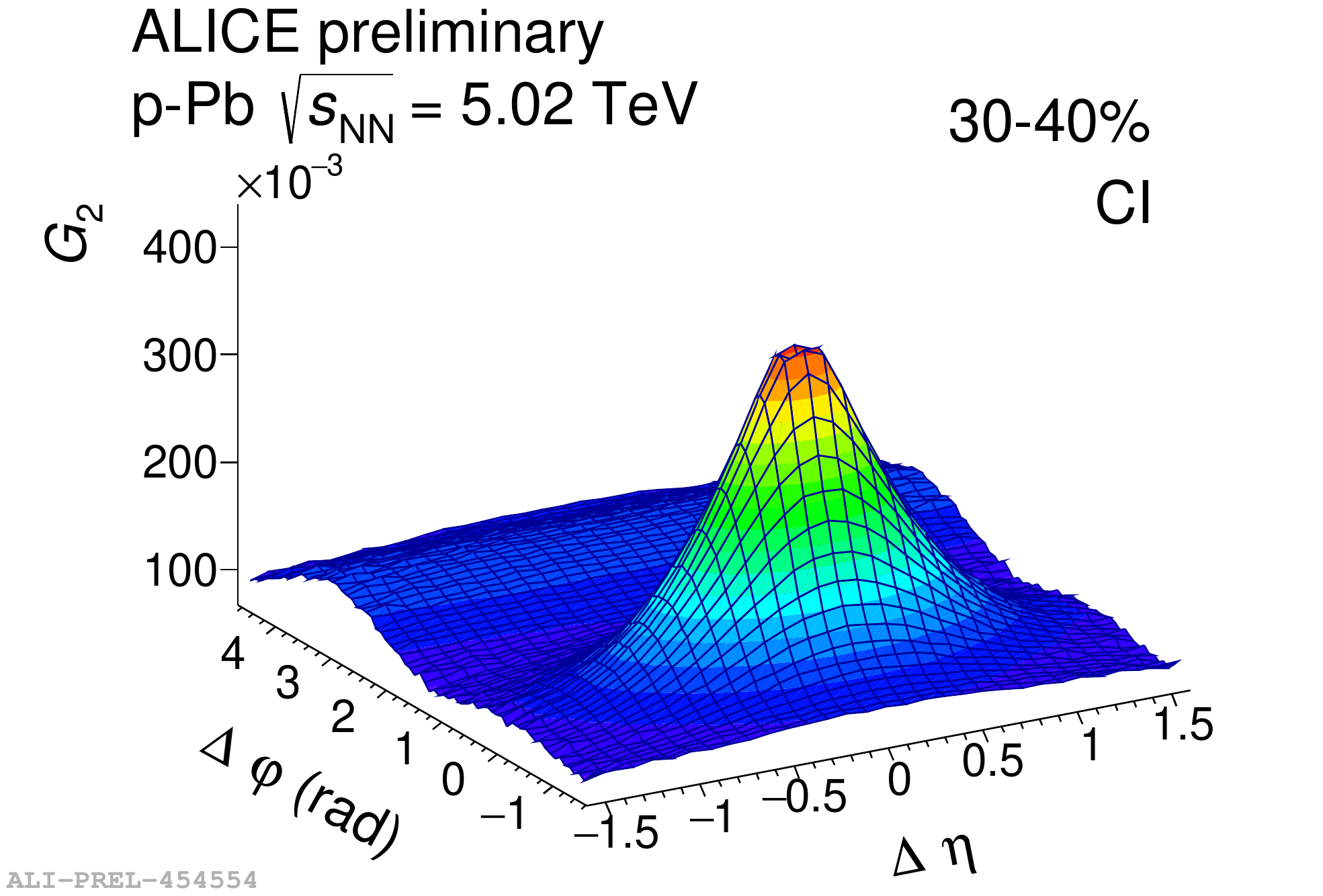}
  \includegraphics[width=0.32\textwidth,keepaspectratio=true,clip=true,trim=0pt 0pt 0pt 0pt]
  {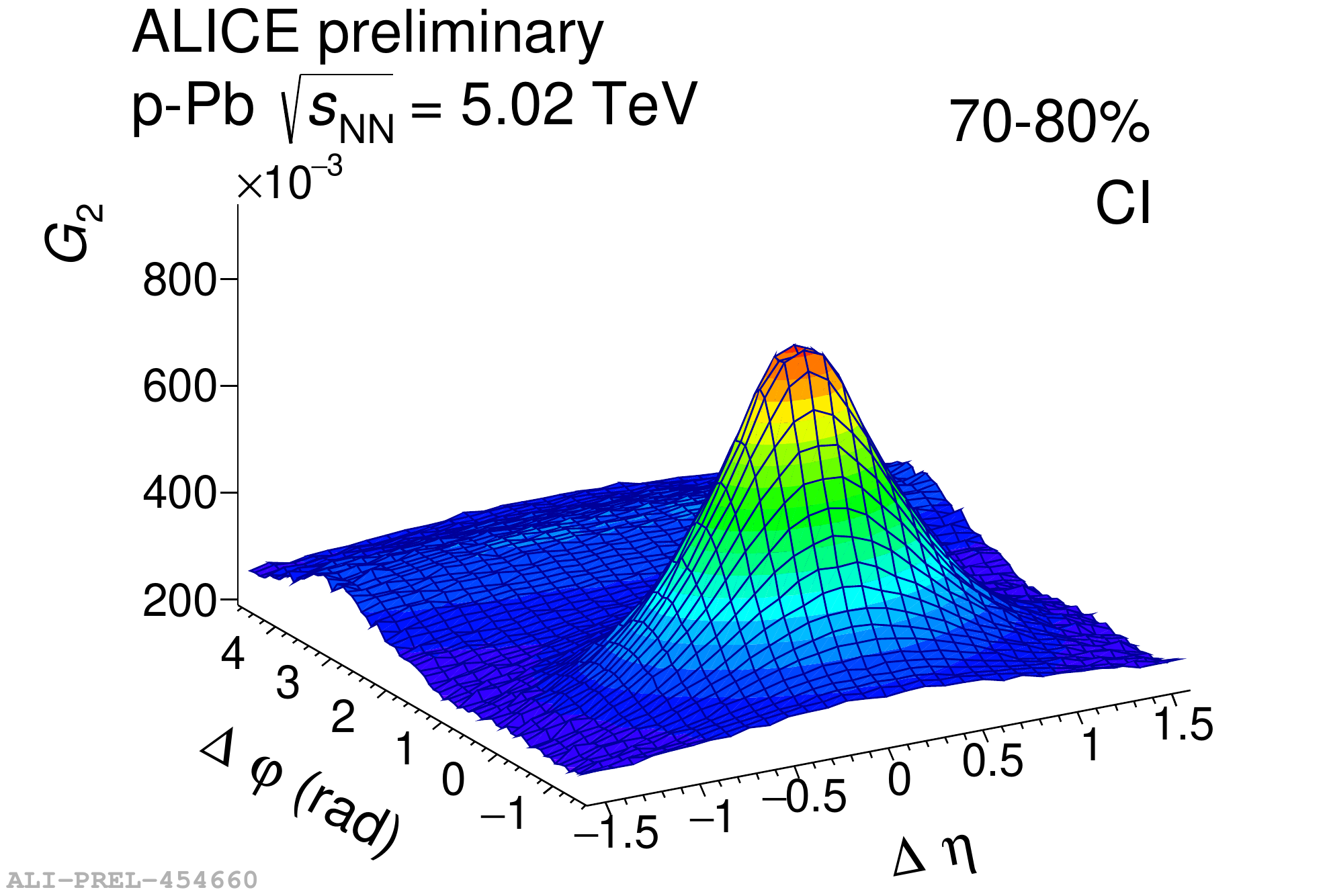}
  \caption{Two-particle transverse momentum correlators $G_{2}^{\rm CD}$ (top row) and $G_{2}^{\rm CI}$ (bottom row) for selected multiplicity classes in p--Pb collisions at $\sqrt{s_{\rm NN}}=5.02\;\text{TeV}$.}
  \label{fig:ppb2dcdci}
\end{figure}
The charge-dependent (${\rm CD}=1/4[(+-)+(-+)-(++)-(--)]$) and charge-independent (${\rm CI}=1/4[(+-)+(-+)+(++)+(--)]$) correlators, $G_{2}^{\rm CD}$ and $G_{2}^{\rm CI}$, respectively, are measured as a function of pair longitudinal, $\Delta\eta$, and azimuthal, $\Delta\varphi$, separation~\cite{ALICE:2019smr}. 
Figures~\ref{fig:pp2dcdci} and~\ref{fig:ppb2dcdci} present $G_{2}^{\rm CD}$ and $G_{2}^{\rm CI}$ correlators for the pp and p--Pb systems, respectively, while Fig.~\ref{fig:ppppbproj} shows their longitudinal and azimuthal projections.
The amplitudes of the correlations decrease monotonically with the collision multiplicity. The CD correlators feature a flat away side associated with the absence of charge-dependent effects, while the CI correlators exhibit a long-range ridge which might be related to momentum conservation. Both correlators show a prominent near side peak, whose longitudinal and azimuthal  widths  are extracted using the same procedure described in Ref.~\cite{ALICE:2019smr}. 
\begin{figure}
\centering
  \includegraphics[width=0.44\textwidth,keepaspectratio=true,clip=true,trim=0pt 0pt 30pt 20pt]
  {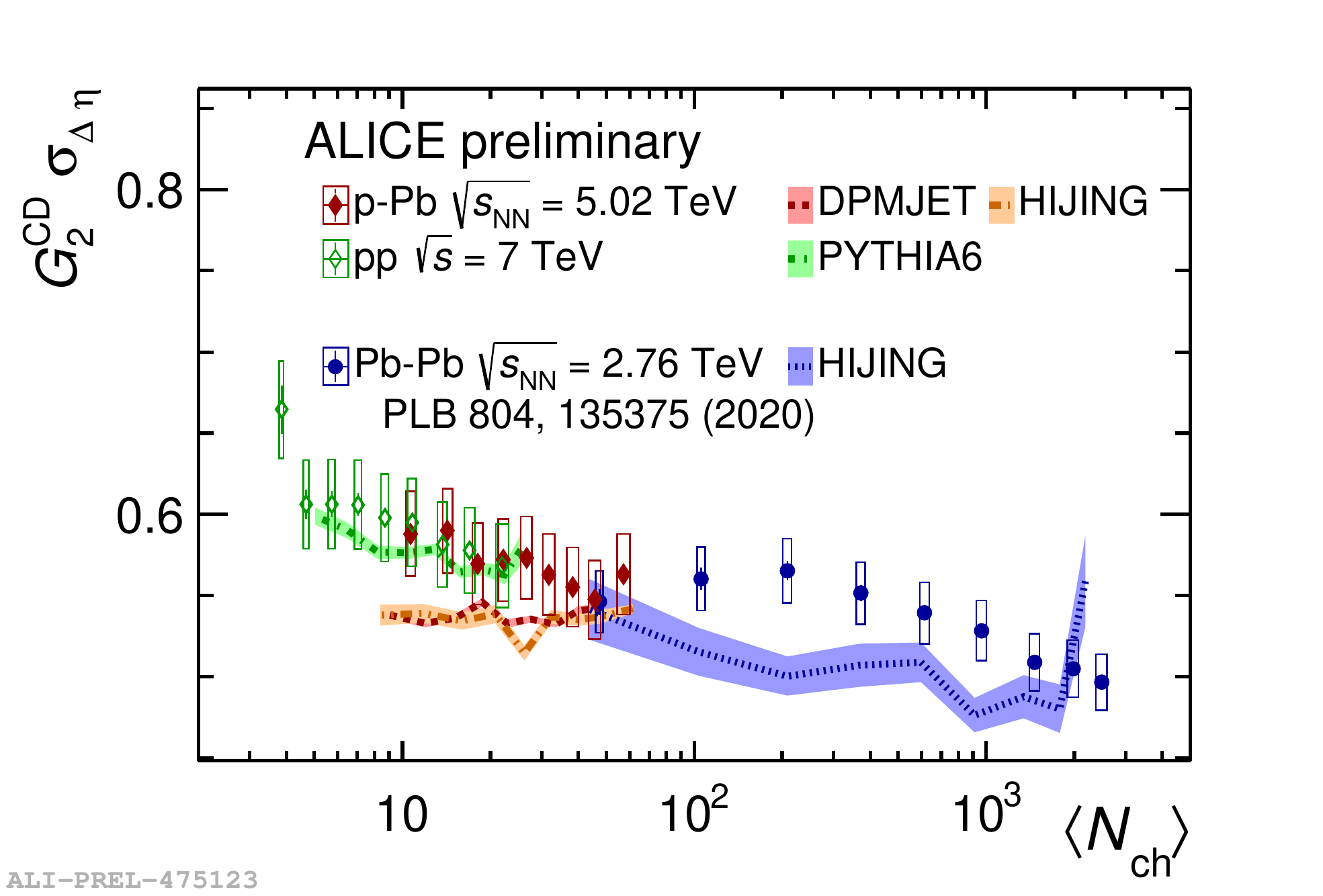}
  \hspace{0.1in}%
  \includegraphics[width=0.44\textwidth,keepaspectratio=true,clip=true,trim=0pt 0pt 30pt 20pt]
  {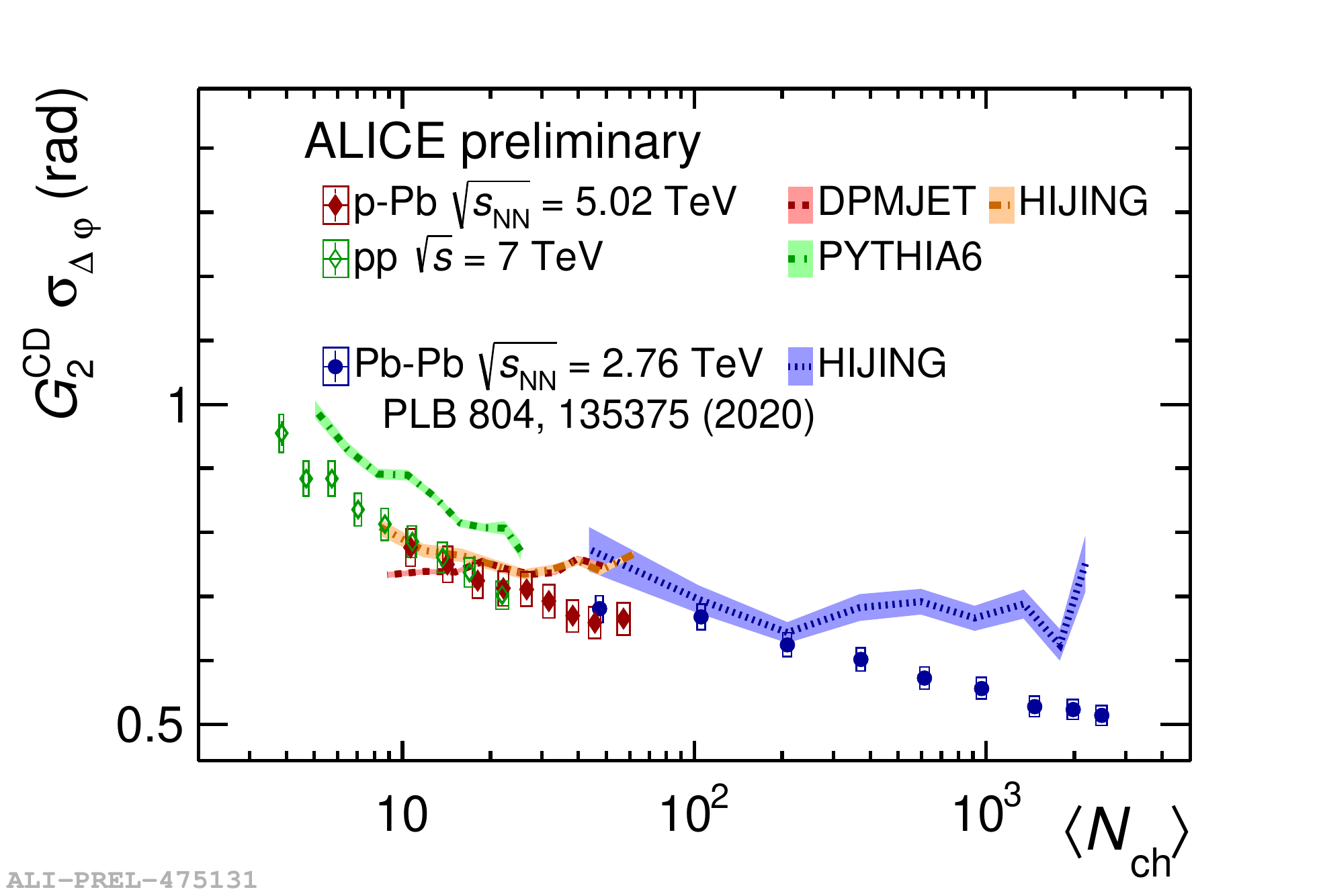}\\
  \includegraphics[width=0.44\textwidth,keepaspectratio=true,clip=true,trim=0pt 0pt 30pt 20pt]
  {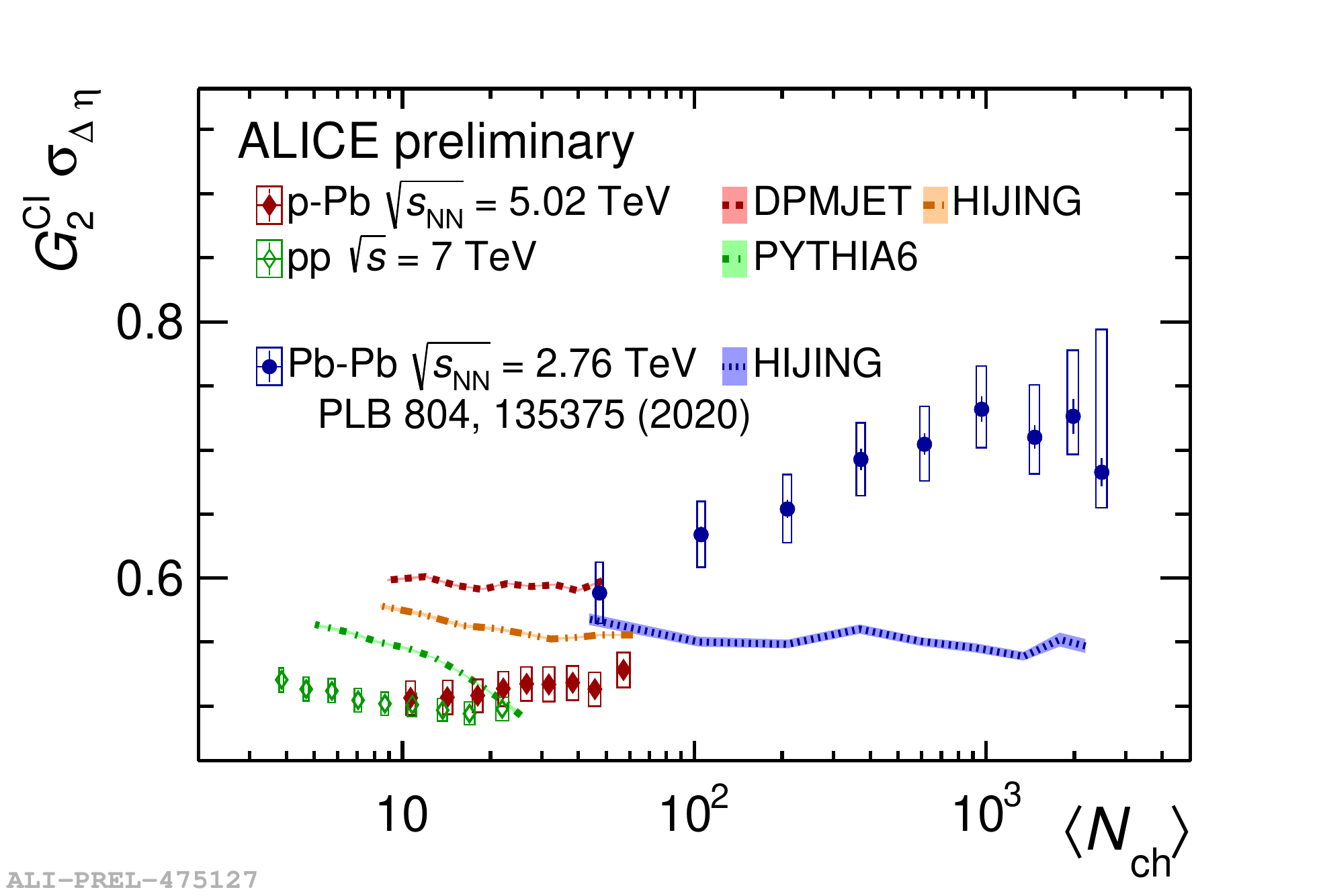}
  \hspace{0.1in}%
  \includegraphics[width=0.44\textwidth,keepaspectratio=true,clip=true,trim=0pt 0pt 30pt 20pt]
  {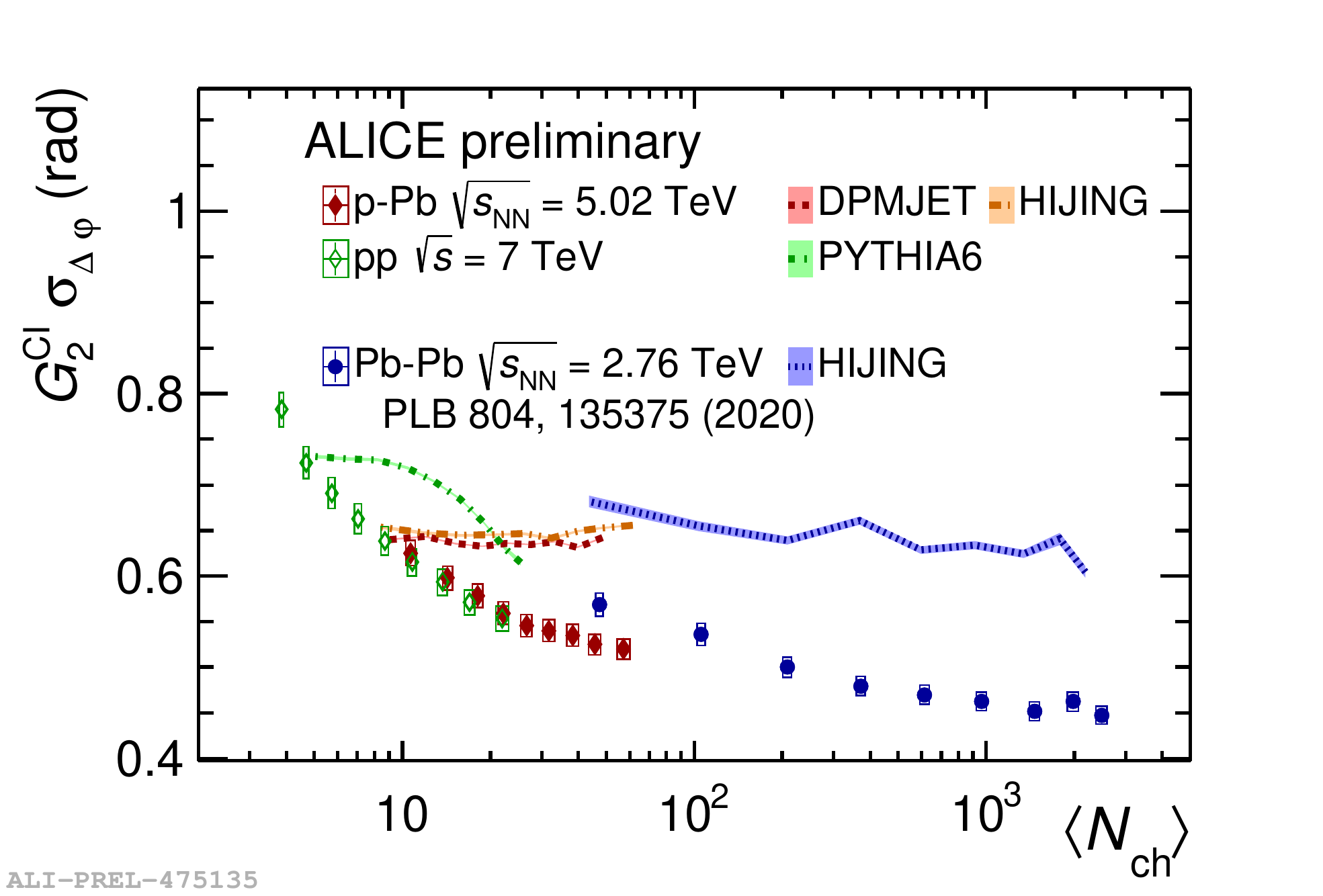}
\caption{Evolution with the average charged particle multiplicity of the longitudinal (left) and azimuthal (right) widths of the two-particle transverse momentum correlations $G_{2}^{\rm CD}$ (top row) and $G_{2}^{\rm CI}$ (bottom row) in pp, p--Pb, and Pb--Pb collisions at $\sqrt{s} = 7\;\text{TeV}$, $\sqrt{s_{\rm NN}} = 5.02\;\text{TeV}$, and $\sqrt{s_{\rm NN}} = 2.76\;\text{TeV}$, respectively, compared to models.}
\label{fig:cdciwidths}
\end{figure}
\section{Results and discussion}
The evolution of the longitudinal and azimuthal widths of the $G_{2}^{\rm CD}$ and $G_{2}^{\rm CI}$ correlators with multiplicity in pp, p--Pb, and Pb--Pb collisions systems is shown in Fig~\ref{fig:cdciwidths} in comparison with the results from the PYTHIA 6 (pp), DPMJET (p--Pb), and HIJING (p--Pb and Pb--Pb) event generators. Azimuthally, both correlators narrow with increasing multiplicity in the three systems, a behaviour reproduced only by PYTHIA 6, although its trend for the CI correlator diverges from that of the data. This azimuthal narrowing is consistent with increasing radial flow and average $p_{\rm T}$ with multiplicity. In the longitudinal direction, the CD correlator shows a narrowing from pp to the low multiplicity Pb--Pb, where the correlator undergoes some broadening that is countered toward higher multiplicities. 
Only PYTHIA is able to describe the behaviour shown by the data. For the CI correlator, the broadening in the Pb--Pb system, interpreted as a fingerprint of viscous effects, is not seen in both the pp and p--Pb systems.
\section{Conclusions}
There is no significant longitudinal broadening of $G_{2}^{\rm CI}$ with increasing multiplicity in pp and p--Pb, in contrast to the broadening measured in Pb--Pb collisions. Thus, based on the longitudinal broadening of $G_{2}^{\rm CI}$, there is no evidence of viscous effects in these systems. This  can also be interpreted as the system being too short-lived for viscous forces to play a significant role.
Only PYTHIA reproduces the behaviour of the CD correlator, while it misses that of the CI correlator.
\bibliographystyle{JHEP}
\bibliography{bibliography.bib}
\end{document}